\documentclass[sigconf]{acmart}

\usepackage{booktabs} 
\usepackage{hyperref}
\usepackage{cleveref}

\usepackage{algorithm}
\usepackage{algorithmic}
\usepackage{amsmath}
\usepackage{amsthm}
\usepackage{balance}
\usepackage[caption=false]{subfig}
\usepackage{extarrows}
\usepackage{graphicx}
\usepackage{multirow}
\usepackage{threeparttable}
\usepackage{tikz}
\usetikzlibrary{fit,positioning,arrows,automata,calc}

\newcounter{ALC@tempcntr}
\newcommand{\LCOMMENT}[1]{%
  \setcounter{ALC@tempcntr}{\arabic{ALC@rem}}
  \setcounter{ALC@rem}{1}
  \item \{#1\}
  \setcounter{ALC@rem}{\arabic{ALC@tempcntr}}
}

\newcommand*\txtcircled[1]{\tikz[baseline=(char.base)]{
            \node[shape=circle,draw,inner sep=0.5pt] (char) {#1};}}

\setcopyright{rightsretained}

\acmISBN{978-1-4503-4896-6/17/06}
\copyrightyear{2017} 
\acmYear{2017} 
\setcopyright{acmlicensed}
\acmConference{WebSci'17}{}{June 25-28, 2017, Troy, NY, USA.}\acmPrice{15.00}\acmDOI{http://dx.doi.org/10.1145/3091478.3091493}
\acmISBN{978-1-4503-4896-6/17/06}

\begin{document}
\title{Hierarchical Change Point Detection on Dynamic Networks}

\author{Yu Wang$^*\quad$ Aniket Chakrabarti$^*\quad$ David Sivakoff$^\#\quad$ Srinivasan Parthasarathy$^*$}
\affiliation{$*$ Department of Computer Science and Engineering, $\quad\#$ Department of Statistics\\
The Ohio State University, Columbus, Ohio, USA\\}
\email{wang.5205@osu.edu,srini@cse.ohio-state.edu}

\renewcommand{\shortauthors}{Y. Wang et al.}

\begin{abstract}
This paper studies change point detection on networks with community structures. 
It proposes a framework that can detect both local and global changes in networks efficiently. 
Importantly, it can clearly distinguish the two types of changes. 
The framework design is generic and as such several state-of-the-art change point detection algorithms can fit in this design.
Experiments on both synthetic and real-world networks show that this framework can accurately detect changes while achieving up to 800X speedup.
\end{abstract}

%
%



\maketitle

\section{Introduction}
Anomaly detection on networks is a problem arising in various areas: 
from intrusion detection~\cite{harshaw2016graphprints} to fraud detection~\cite{eberle2016identifying}, from email network~\cite{peel2014detecting} to fMRI image~\cite{koutra2016deltacon}. 
One problem of particular interest is change point detection on dynamic social networks \cite{AkogluTK14,ranshous2015anomaly}. 
Social networks are known to have the hierarchical structure, 
where the most well-known one is the community structure~\cite{fortunato2010community,ruan2015community}. 
Similar nodes are densely connected and form a community, 
while dissimilar nodes reside in different communities and are less likely to be connected. 
The interactions (or edges) among nodes can be classified as inter-community interactions and intra-community interactions. 
The intra-community interactions are more likely to be present (in unweighted networks) or have more weight (in weighted networks) than the inter-community interactions. 

\begin{figure}[!bth]
\hspace{-0.0in}
\begin{minipage}{240pt}
\includegraphics[width=240pt,height=140pt]{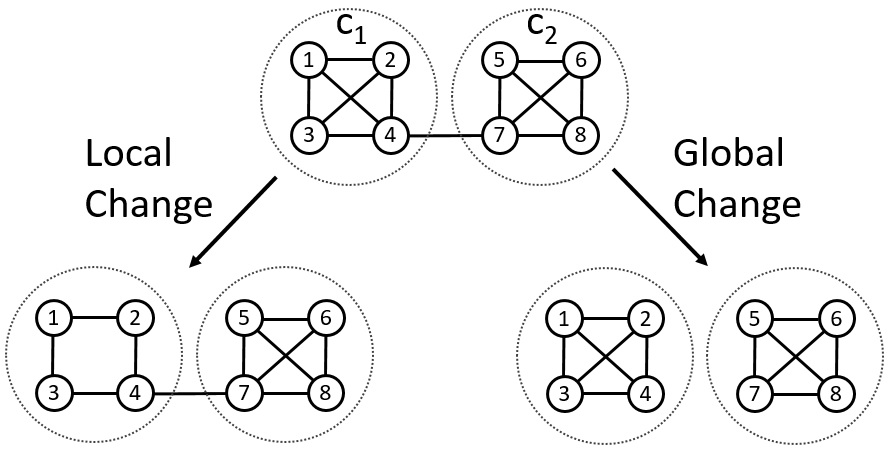}%
\captionsetup{width=240pt}
\vspace{-0.7cm}
\caption{\small{Toy example to distinguish global change and local change on a two-community graph. 
The edges removal within $c_1$ has only local impact, and hence is a local change; 
the edge removal across $c_1$ and $c_2$ has global impact (to disconnect the graph), and hence is a global change.}}
\label{fig:ToyExpGlobalLocal}
\end{minipage}
\end{figure}

Most state-of-the-art change point detection approaches 
\cite{koutra2016deltacon,la2014anomaly,harshaw2016graphprints,eberle2016identifying,wang2017ijcai} 
do not consider the \emph{hierarchical structure} in a network, but treat the network structure as flat.  
Hence they can hardly distinguish changes in those two kinds of interactions (intra and inter). 
If there is an event associated with a particular community, 
the change happens within that community only, 
and may not affect the global network too much due to its locality and small scale. (Toy example in~\Cref{fig:ToyExpGlobalLocal})

For example, consider the email network of a very big computer science department. 
If a specific lab is approaching a paper submission deadline, 
the frequency of email exchange within this lab might outburst. 
Since the whole department is very big and the daily email throughput of the whole department is high, 
it is unlikely to detect the change associated with that lab at the global level using the aforementioned algorithms. 

On the other hand, if a single community is significantly larger than the others, 
a significant change in that community will dominate over the other smaller communities. 
Therefore, regardless of whether there is any change to the rest of the network, 
the aforementioned algorithms will classify this as a change point.
That is, the above algorithms do not answer where the change comes from, 
neither can they answer if the change comes from within a huge community or from the inter-community interaction. 

Continuing with the previous email network example, 
if a lab consists of half the people in the department (either a big group or a small department), 
then the outburst of emails within this lab is highly likely to be detected as a global event. 
Meanwhile, if several groups are jointly working on a funding proposal, 
it will lead to the email burst between communities, 
and can also be detected as a global event. 
The above algorithms can not distinguish between these two global events. 
Here we see that the existing algorithms either miss minor and local changes, 
or fail to distinguish between strong local changes and global changes. 
A recent work (\cite{lagenerating}) can answer which pairs of nodes result in change, 
but it fails to account for the community structure. 

Intuitively, the intra-community interaction captures local information while the inter-community interaction captures global information. 
Hence, we expect to see the former one evolves differently from the global evolution pattern while the latter one should be similar to the global evolution pattern \cite{asur2009event}.

In this paper, we show that the network evolution pattern can be decomposed into inter-community evolution and intra-community evolution. 
We show that running change point detection algorithms at the global level could miss changes associated with a community, 
and that change can be detected when we run the algorithms within that community. 
We also show that the inter-community interaction can be approximated by a hyper-network in which each hyper-node is a contracted community \cite{blondel2008fast}. 
And this contraction also gives us computational benefits. 

This paper is organized as follow: we first review the related work, then describe the proposed framework. 
After that, we introduce the experiments and provide the analysis. 
Finally, we conclude the paper and discuss future works.

\section{Related Work}
There are two recent surveys~\cite{AkogluTK14,ranshous2015anomaly} on change point detection. 
Most state-of-the-art works~\cite{koutra2016deltacon,wang2017ijcai} do not consider the hierarchical structure in a network; 
Other works~\cite{moreno2013network,peel2014detecting,bridges2015multi} mention the hierarchy in their papers, 
and they all make specific assumption about the underlying generative model: 
Moreno's work~\cite{moreno2013network} assumes a network is generated from a mixed Kronecker product graph model (mKPGM), 
which is generated recursively from a seed matrix. 
The block of the matrix at each level resembles the community. 
Peel's work~\cite{peel2014detecting} assumes a network is generated from a generalized hierarchical random graph model (GHRG), 
which organizes the network as a tree of which the leaves are nodes in the network and the internal nodes are communities. 
Bridges's work~\cite{bridges2015multi} assumes a network is generated from the generalized block two-level Erd{\H{o}}s-R{\'e}nyi (GBTER) model, 
which posits the inter-community and intra-community edges are formed in different ways. 
Although they mention the hierarchical structure, 
they fail to distinguish between the change associated to a community and the change associated to the whole network.
We design a framework that provides a systematic way to detect and distinguish between local and global changes. 
Our framework is generic and supports several state-of-the-art change point detection algorithms. 
Another drawback of these network hypothesis test based approaches is their low efficiency due to bootstrapping: 
Peel's work does not scale very well in our experiments, 
and its efficiency is improved up to 60 times when superimposed by our framework. 

Community detection~\cite{fortunato2010community} is also a time-consuming task. 
The Louvain method~\cite{blondel2008fast} alternatively contracts and detects communities on the networks, 
and has been one of the most efficient community detection algorithm for several years. 
Its efficiency comes from the network size reduction via contracting. 
We use the same idea for efficient change point detection.

Another recent work (\cite{lagenerating}) considers attributing the global change score to node pairs. 
It defines change scores at edge level, flags out the top ranked edges and their neighbors as changed regions. 
That work does not consider the community structure, and its focus is on visualization. 
Our work distinguishes between inter-community interaction and intra-community interaction, 
and shows the former one can approximate the global evolution. 
Moreover, our framework reduces the running time of the state-of-the-art algorithms~\cite{peel2014detecting,koutra2016deltacon,wang2017ijcai}.

\section{Methodology}
\begin{figure}[!htb]
\begin{tikzpicture}
\tikzstyle{main}=[circle, minimum size = 5mm, thick, draw =black!80, node distance = 10mm]
\tikzstyle{connect}=[-latex, thick]
\tikzstyle{box}=[rectangle, draw=black!100]
  \node[box,draw=white!100] (Latent) {\textbf{Latent}};
  \node[main] (M1) [right=0.5cm of Latent] {$L_1$};
  \node[main] (M2) [right=of M1] {$L_2$};
  \node[main] (M3) [right=of M2] {$L_3$};
  \node[main] (Mt) [right=of M3] {$L_t$};
  \node[main,fill=black!10] (G1) [below=of M1] {$G_1$};
  \node[main,fill=black!10] (G2) [right=of G1,below=of M2] {$G_2$};
  \node[main,fill=black!10] (G3) [right=of G2,below=of M3] {$G_3$};
  \node[main,fill=black!10] (Gt) [right=of G3,below=of Mt] {$G_t$};
  \node[box,draw=white!100] (Observed) [left=0.5cm of G1] {\textbf{Observed}};
  \path (M3) -- node[auto=false]{\ldots} (Mt);
  \path (G1) edge [connect] (G2)
        (G2) edge [connect] (G3)
        (G3) -- node[auto=false]{\ldots} (Gt);
  \path (M1) edge [connect] (G1);
  \path (M2) edge [connect] (G2);
  \path (M3) edge [connect] (G3);
  \path (Mt) edge [connect] (Gt);

  \draw [dashed, shorten >=-1cm, shorten <=-1cm]
    ($(Latent)!0.5!(Observed)$) coordinate (a) -- ($(Mt)!(a)!(Gt)$);

\end{tikzpicture}
\caption{\small Representation of the underlying generative process. $L_t$s are latent generative models, and $G_t$s are observed snapshots. $G_t$ is assumed to be a sample (generated) from $L_t$. ``Significant'' change points are $\{t\mid L_t\neq L_{t-1}\}$. Note that graph snapshots and latent models can represent both local structure as well as global structure.}
\label{fig:graphicalModel}
\end{figure}
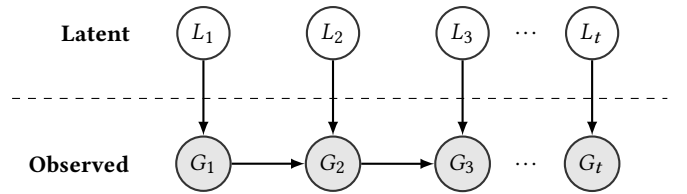

We follow the Type 4, ``Event and Change Detection'', in \cite{ranshous2015anomaly}, 
which is to find snapshots from a network snapshot sequence that are ``significantly'' different from their predecessors. Like~\cite{wang2017ijcai}, we also assume each snapshot is generated from a generative model. Snapshots from the same generative model resemble each other, although they are not identical due to randomness. The change of the generative model leads to ``significant change'' of the snapshots. This assumption can be illustrated as a Markov network (\Cref{fig:graphicalModel}).

Our work differs from~\cite{wang2017ijcai} in that we study networks with the community structure, 
where each snapshot observation consists of multiple components, corresponding to communities and their interactions. 
In \Cref{fig:graphicalModel}, the generative models are latent and we can only observe the generated snapshots. 
Existing approaches have two directions: to estimate the generative model from multiple snapshots and then compare the estimated model~\cite{peel2014detecting,bridges2015multi,wang2017ijcai}; 
or to compute the difference of the snapshots directly~\cite{koutra2016deltacon}. 
Our framework supports both types of approaches. 
In both cases, the problem and the solution can be generalized as~\cite{ranshous2015anomaly}: 
given a network snapshot sequence $\{G_t\}$ and a dissimilarity function (outlier score) $f:\{G_t\}\times\{G_t\}\to\mathcal{R}$, 
a change is defined as a time point $t$, 
such that $f(G_t,G_{t-1})$$>$$th_0$, where $th_0$ is some pre-specified threshold. 
Change point detection is to find all time stamps $t$s at which changes occur. 
All the aforementioned works run on the original network, 
and hence detect changes at the global level. 

We define intra-community change and inter-community change in a similar fashion: 
let $C_{it}$ be the community $i$ at time $t$, 
the intra-community change point is defined as a time point $t$ such that $f_l(C_{it},C_{i,t-1})>th_1$.
By contracting the community $C_{it}$ in the network $G_t$ into a hyper node $n_i$, 
one gets a contracted network $G_t^c$. 
The inter-community change point is defined as a time point $t$ such that $f_c(G_{t}^c,G_{t-1}^c)>th_2$.
Our goal is to find both the intra-community change (local change) as well as the inter-community change (global change). 

\noindent\textbf{Problem Definition} Given a network sequence $\{G_t\}_{t=1}^T$, its community assignment $\{th_i\}_{i=1}^k$, 
and dissimilarity functions $f_l(\cdot,\cdot),f_c(\cdot,\cdot)$, find all the time points $t\in\{2,\cdots,T\}$ such that either 
1) $f_l(C_{it},C_{i,t-1})>th_1$, or 2) $f_c(G_{t}^c,G_{t-1}^c)$$>$$th_2$ for some pre-specified threshold $th_1,th_2$.

\subsection{Framework Description}
\label{subsect:framework}
\begin{figure}[!bth]
\hspace{-0.0in}
\begin{minipage}{240pt}
\includegraphics[width=240pt,height=180pt]{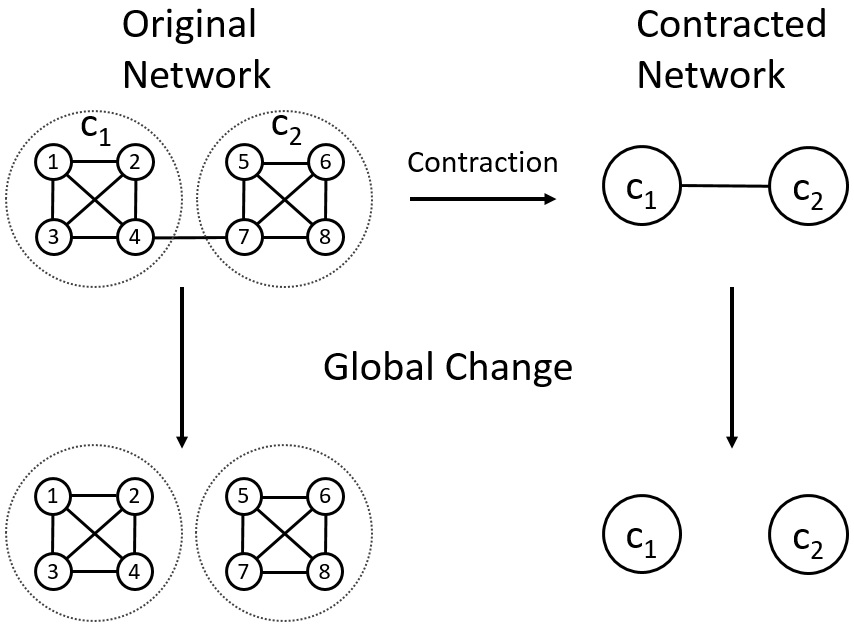}%
\captionsetup{width=240pt}
\vspace{-0.7cm}
\caption{\small{Toy example to illustrate contraction on the previous two-community graph. 
Communities $c_1,c_2$ are contracted into hyper-node $c_1,c_2$ respectively. 
Global change in the original network (graph disconnection) is captured by the change in the contracted network.
}}
\label{fig:ToyExpContraction}
\end{minipage}
\end{figure}

We propose the following framework (\Cref{Algo1}) which reveals both the local change as well as the global change. 
In Line 1, we first partition communities on the first snapshot. 
We assume the community assignment does not change over time. 
If there really is a severe change in community assignment, 
the event will be detected as a global change. 
Between global changes, the community assignment will remain constant. Hence one can apply the local change detection part of the framework to the subsequences in which community assignment remains similar. 
One may argue that it is more robust to partition communities on the aggregated network from the first couple of snapshots. 
Our reasoning is that if no global change happens during the first couple of snapshots, 
the community assignment/structure should remain the same; 
otherwise we flag out that snapshot as abnormal (to be investigated), 
and redo the community partition on the new snapshot. 
The for-loop from Line 2 to Line 12 scans over all the snapshots. 
The original snapshot is contracted in Line 3 into a weighted (hyper-)network in which each hyper-node corresponds to a community in the original snapshot (\Cref{fig:ToyExpContraction}). 
If the original network is unweighted, the weight of the hyper-edge is the number of actual inter-community edges divided by the number of all possible inter-community edges. 
Otherwise the weight of the hyper-edge is the summation of the inter-community edge weights divided by the sizes of the two communities. 
The contraction operation is widely used in hierarchical community partition algorithms~\cite{karypis1998multilevelk,fortunato2010community}. 
The for-loop from Line 4 to Line 8 iterates over all the communities. 
The provided change point detection algorithm runs on the community level, and returns the dissimilarity score between two consecutive (sub-)networks. 
The branch in Line 9 detects the global change by running the detection algorithm on the contracted (hyper-) network. 

\begin{algorithm}[!htb]
\caption{Hierarchical Change Point Detection Framework}
\label{Algo1}
\begin{algorithmic}[1]
\begin{small}
\REQUIRE network sequence $\{G_t\}_1^t$, any change point detection algorithm \textit{IsChanged($\cdot$,$\cdot$)}
\ENSURE global change points $changeSet$, local change points $changeSet_i$ in community $i$ 
\STATE $\{c_i\}_1^k=$ partition$(G_1)$; \COMMENT{community partition}
\FOR {$t=2$ to $T$}
  \STATE $G_t^c=$contract$(\{c_i\}_1^k)$;
  \LCOMMENT{Below: local change detection}
  \FOR {$i=1$ to $k$}
    \IF {\textit{IsChanged}$(c_{it},c_{i,t-1})$}
      \STATE $changeSet_i=changeSet_i\cup\{t\}$;
    \ENDIF
  \ENDFOR
  \LCOMMENT{Below: global change detection}
  \IF {\textit{IsChanged}$(G_t^c,G_{t-1}^c)$}
    \STATE $changeSet=changeSet\cup\{t\}$;
  \ENDIF
\ENDFOR
\end{small}
\end{algorithmic}
\end{algorithm}

Community partitioning gives us efficacy while network contraction gives us efficiency: 
each community is a fine-grained sub-network, 
and running detection algorithm on each community can avoid fluctuation and noises from neighboring communities, 
and hence the detected change points are solely associated with this particular community; 
contraction significantly reduces network size, 
and in turn, reduces the running time of the algorithms whose complexity is proportional to the network size. 
On the other hand, network size reduction also implies information loss, 
and therefore the quality of change points detected on the contracted network is inferior to that on the original network. 
Fortunately, our experiments show that the ranking of outlier scores on the contracted network is similar to that on the original network, 
which implies that the inter-community evolution can indeed capture global evolution pattern. 

\subsection{Algorithms}
We superimpose the framework on three state-of-the-art algorithms~\cite{koutra2016deltacon,peel2014detecting,wang2017ijcai}, 
\subsubsection{DeltaCon}
DeltaCon \cite{koutra2016deltacon} computes the outlier score from the snapshots directly. 
It extracts a feature vector from each snapshot using personalized PageRank, 
and computes the rooted Euclidean distance of two consecutive feature vectors. \vspace{-0.05in}
\subsubsection{LetoChange}
LetoChange \cite{peel2014detecting} assumes the network is generated from GHRG model in which the network has a tree-like structure: leaves correspond to nodes while internal tree nodes correspond to communities. 
The network sequence is partitioned into equal sized sliding windows, 
and the model parameters are estimated within a window. 
The outlier is flagged out if the model parameters estimated change greater than a threshold. 
It bootstraps networks from the generative model to calculate a proper threshold. 
To sample multiple networks from a generative model is time-consuming. 
We bootstrap outlier scores for threshold determination, which gains much computational benefit (detailed at the end of this section). \vspace{-0.05in}
\subsubsection{EdgeMonitoring}
EdgeMonitoring~\cite{wang2017ijcai} follows LetoCh-ange's fashion: first estimate the model parameters from a window (several snapshots), 
then detect change points based on model parameters comparison. 
EdgeMonitoring estimates the edge probability directly from each window (for the unweighted network) 
and calculates the Kullback-Leibler divergence of the sequences of two consecutive windows. 
We modify it to account for weighted networks: we normalize the weights at each snapshot, 
and average over snapshots within a window. 
This averaged weight is used as the probability estimate in the above KL divergence calculation. 

\begin{algorithm}[!thb]
\caption{Hierarchical Change Point Detection Framework with Threshold Determination}
\label{Algo2}
\begin{algorithmic}[1]
\begin{small}
\REQUIRE network sequence $\{G_t\}_1^t$, any change point detection algorithm \textit{IsChanged($\cdot$,$\cdot$)}
\ENSURE global change points $changeSet$, local change points $changeSet_i$ in community $i$ 
\STATE $\{c_i\}_1^k=$ partition$(G_1)$; \COMMENT{community partition}
\FOR {$t=2$ to $T$}
  \STATE $G_t^c=$contract$(\{c_i\}_1^k)$;
  \LCOMMENT{Below: local change detection}
  \FOR {$i=1$ to $k$}
    \STATE $OutlierScore_{i,t}=$ \textit{IsChanged}$(c_{it},c_{i,t-1})$
  \ENDFOR
  \LCOMMENT{Below: global change detection}
  \STATE $OutlierScore_t=$ \textit{IsChanged}$(G_t^c,G_{t-1}^c)$
\ENDFOR
\FOR {$i=1$ to $k$}
  \STATE $Threshold_i=$ Bootstrap$(OutlierScore_i).$percentile$(0.95)$
  \STATE $changeSet_i=\{t\mid OutlierScore_{i,t}>Threshold_i\}$
\ENDFOR
\STATE $Threshold=$ Bootstrap$(OutlierScore).$percentile$(0.95)$
\STATE $changeSet=\{t\mid OutlierScore_t>Threshold\}$
\end{small}
\end{algorithmic}
\end{algorithm}

Although DeltaCon and LetoChange have their own threshold determination strategies, 
the former one's normality assumption does not hold while the latter one's network bootstrap is too time-consuming. 
We use an efficient, permutation test based strategy. 
All the three algorithms use the same way to determine outlier scores: 
first, run the algorithm on the whole snapshot sequence and obtain an outlier score sequence; 
then, use bootstrap to compute the 0.95 confidence level of the outlier scores. 
The rationale of this threshold determination is essentially the same as that of the permutation test~\cite{pitman1937significance}. 
The modified framework is displayed in \Cref{Algo2}, in which the \textit{IsChanged}$(,)$ function again returns the dissimilar score. 

\section{Experiments and Evaluation}
\subsection{Data}
We apply the framework to two synthetic networks and one real world network. 
For the synthetic networks, we test our framework on the Stochastic Block Model (SBM~\cite{karrer2011stochastic}) 
and the Block Two-Level Erd\H{o}s-R{\'e}nyi (BTER~\cite{seshadhri2012community}) model.
SBM assumes edge probability a function of the community membership of the two incident nodes; 
BTER model assumes the intra-community edges are sampled from Erd\H{o}s-R{\'e}nyi model~\cite{erd6s1960evolution}, 
while the inter-community edges are sampled from Chung-Lu model~\cite{chung2002average}. 
For both experiments, we do not change community membership (as discussed in \Cref{subsect:framework}), 
and inject either global or local changes at different time stamps (\Cref{table:event}).

\begin{table}[!ht]
\small
\caption{\small{Changes Injected to the Synthetic Networks}}
\label{table:event}
\begin{threeparttable}
\begin{tabular}{|p{22pt}|p{19pt}|p{80pt}|p{80pt}|}
\hline
Event Order & Time Stamp & SBM (1k nodes, 8 communities) & BTER (100 nodes, 5 communities)\\
\hline
\txtcircled{1} & 16 & connection rate in the largest community $c_0$ reduced by 1/3 (Local Change) & connection rate in the largest community $c_0$ reduced by 1/3 (Local Change) \\
\txtcircled{2} & 31 & inter-community connection rates among $c_0,c_1,c_5,c_6$ reduced by 1/3 (Global) & Chung-Lu sequence in $c_0$ change (Global) \\
\txtcircled{3} & 51 & connection rates in two smallest communities $c_6,c_7$ doubled (Local) & connection rates in two smallest communities $c_3,c_4$ doubled (Local) \\
\txtcircled{4} & 76 & SBM matrix regenerated (Global) & Chung-Lu sequence of all the nodes regenerated (Global)\\
\hline
\end{tabular}
\end{threeparttable}
\end{table}

The SBM experiment has 1k nodes and 8 communities ranging from size 50 to size 300. 
The edge probability is determined by a symmetric stochastic block matrix whose entries are sampled from 
\textit{Uniform}$(0,1)$ with the restriction that main diagonal entries greater than off-diagonal entries.
The BTER experiment has 100 nodes and 5 communities whose sizes are sampled from power a law distribution~\cite{yang2012community}, 
and range from 15 to 25. ER probability is sampled from a beta distribution~\cite{bridges2015multi}, 
while CL sequence is sampled from another power law distribution~\cite{faloutsos1999power}.
Each snapshot is a weighted network where the edge weight is a sample from a binomial distribution and normalized to the range [0,1]. 
Since LetoChange does not support weighted network, we unweight each network by 
retaining the edges with probability equal to the edge weight. 

For the real world network, we use the international trade network\footnote{\footnotesize{www.worldbank.org}} from the year 2001 to the year 2014. 
We select 90 countries/economies and represent them as nodes. 
The network is directed and weighted, 
where the direction specifies import/export and the weight is the trade volume. 
The network is near a clique since almost every two countries have bilateral trade between them.

There are six ground truth communities in the network, corresponding to free trade zone and/or geopolitics alike nations: 
Asia-Pacific Economic Cooperation (APEC), black sea economic cooperation (BSEC), fifteen developed western European countries (EU15), Latin American Integration Association (LAIA) and Southern African Development Community (SADC), Organization of Islamic Cooperation (OIC). 
The Louvain~\cite{blondel2008fast} community partition algorithm is also applied to this network and returns consistent partitions. 

\begin{figure}[!ht]
\begin{minipage}{240pt}
\subfloat[SBM Global (1k nodes, re-scalded and shifted for visualization). Local change \textcircled{3} in the two smallest communities are insignificant at global level, and hence is ignored by both algorithms.\label{fig:SBMGlobalOriginal}]{\includegraphics[width=240pt,height=100pt]{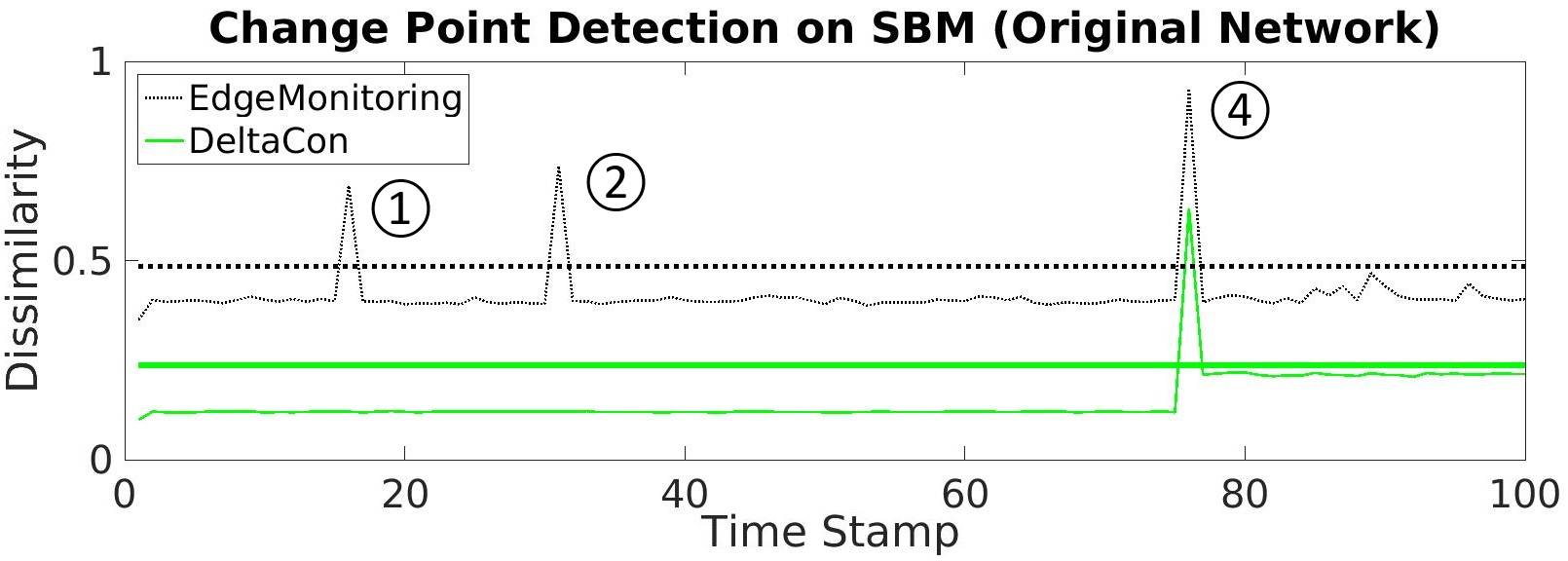}}\hfill
\subfloat[SBM Contract (8 hyper-nodes). EdgeMonitoring (black, top) detects two global events \textcircled{2} and \textcircled{4}, while DeltaCon (green, middle) and LetoChange (red, bottom) detect only one.\label{fig:SBMGlobalContract}]{\includegraphics[width=240pt,height=100pt]{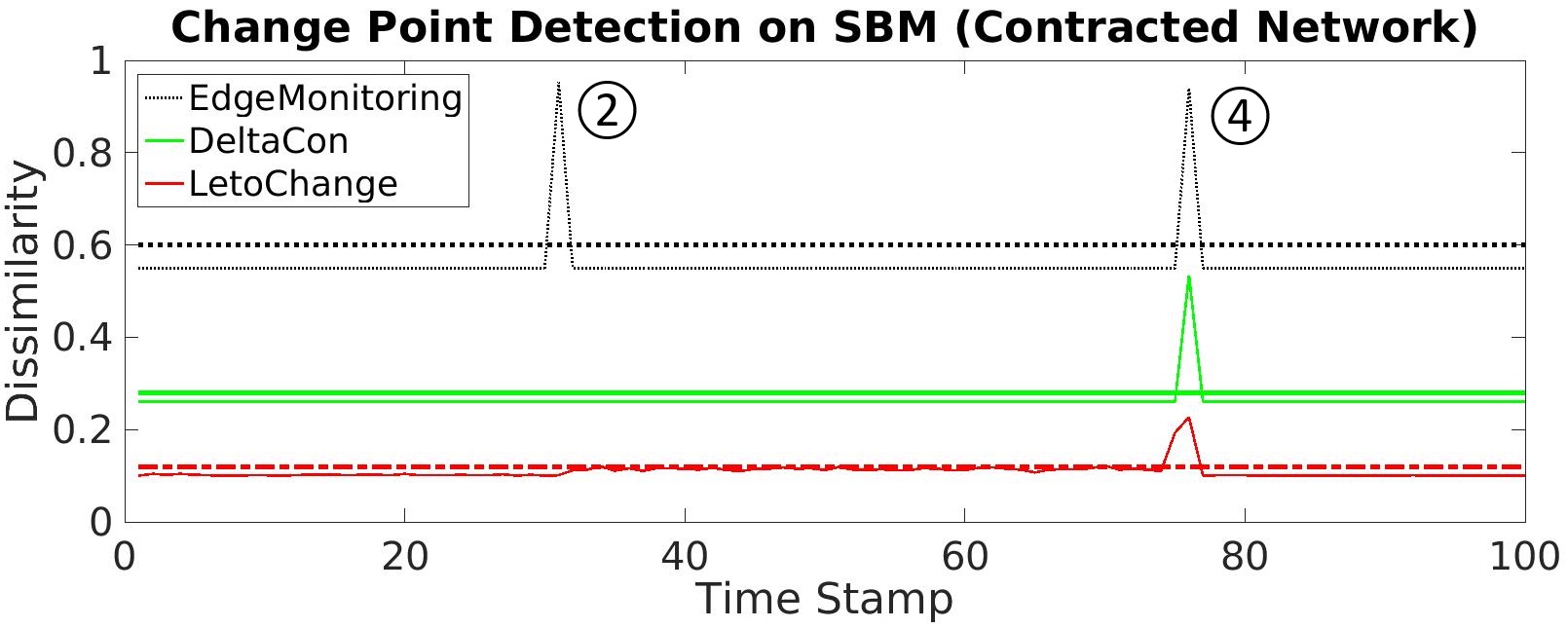}}\hfill
\vspace{-0.1cm}
\end{minipage}
\captionsetup{width=240pt}
\caption{\small{Change point detection on SBM at global level.}}
\label{fig:SBMGlobal}
\end{figure}

\subsection{Experiments and Results}
\subsubsection{Synthetic Network}
We apply our framework with the three aforementioned detection algorithms on the synthetic networks. 
The generation process is described in the previous subsection. 
We present both quality and time efficiency of our framework. 
We report time efficiency in~\Cref{table:efficiency}, and use both qualitative and quantitative methods to validate the quality. 

\Cref{fig:SBMGlobal} shows the change points detected on the SBM network. 
\Cref{fig:SBMGlobalOriginal} is the result on the global (original) network, 
and \Cref{fig:SBMGlobalContract} is the result on the contracted network; 
the curves are outlier scores for various change point detection algorithms and the horizontal bars are corresponding thresholds. 
LetoChange can not finish within 48 hours (\Cref{table:efficiency}) on the original network, 
and hence is not displayed in \Cref{fig:SBMGlobalOriginal}. 
We can see that EdgeMonitoring correctly captures two global changes and a local change in the largest community of the original network (\Cref{fig:SBMGlobalOriginal}). 
After contraction (\Cref{fig:SBMGlobalContract}), EdgeMonitoring only captures the two global changes. 
This is what we can expect since inter-community interaction reflects the global pattern, and contraction only preserves this global pattern. 
Although DeltaCon does not capture both the global changes, 
the quality of DeltaCon on the contracted network is the same as on the original network while achieving 830 times efficiency improvement! 
Note that the quality of change point detection on the original network is totally determined by the detection algorithm itself. 
Applying our framework (i.e. to contract the network) does not deteriorate the quality. 
A similar observation holds for LetoChange. 

We note that it is the outlier score ranking (say, top $\alpha_s\%$ as in~\cite{wang2017ijcai}), 
rather than the absolute values, that matters in change point determination. 
Hence we use the Normalized Discounted Cumulative Gain (NDCG~\cite{wang2013theoretical}) 
scores to evaluate how good the contracted one approximates the original one. 
NDCG score quantifies the difference between two rankings, and perfect match results in $\text{NDCG}=1.0$. 
\[\text{NDCG}_T=\frac{\text{DCG}_T}{\text{IDCG}_T},\qquad\text{where}\]

\[\text{DCG}_T=\sum_{i=1}^T\frac{2^{r_i}-1}{\log_2(i+1)},\quad\text{IDCG}_T=\sum_{i=1}^T\frac{2^i-1}{\log_2(i+1)}\]
DCG is calculated on the contracted network, while IDCG (reference) is calculated on the original network. 
The outlier scores from two networks are sorted in descending order respectively. 
$r_i$ is the number of matches between the target ranking and the reference ranking within first $i$ scores, and $T$ is the number of snapshots.
The NDCG scores for EdgeMonitoring and DeltaCon are $0.832$, $0.666$ respectively, indicating reasonably good approximation. 

\begin{figure}[!ht]
\begin{minipage}{240pt}
\includegraphics[width=240pt,height=100pt]{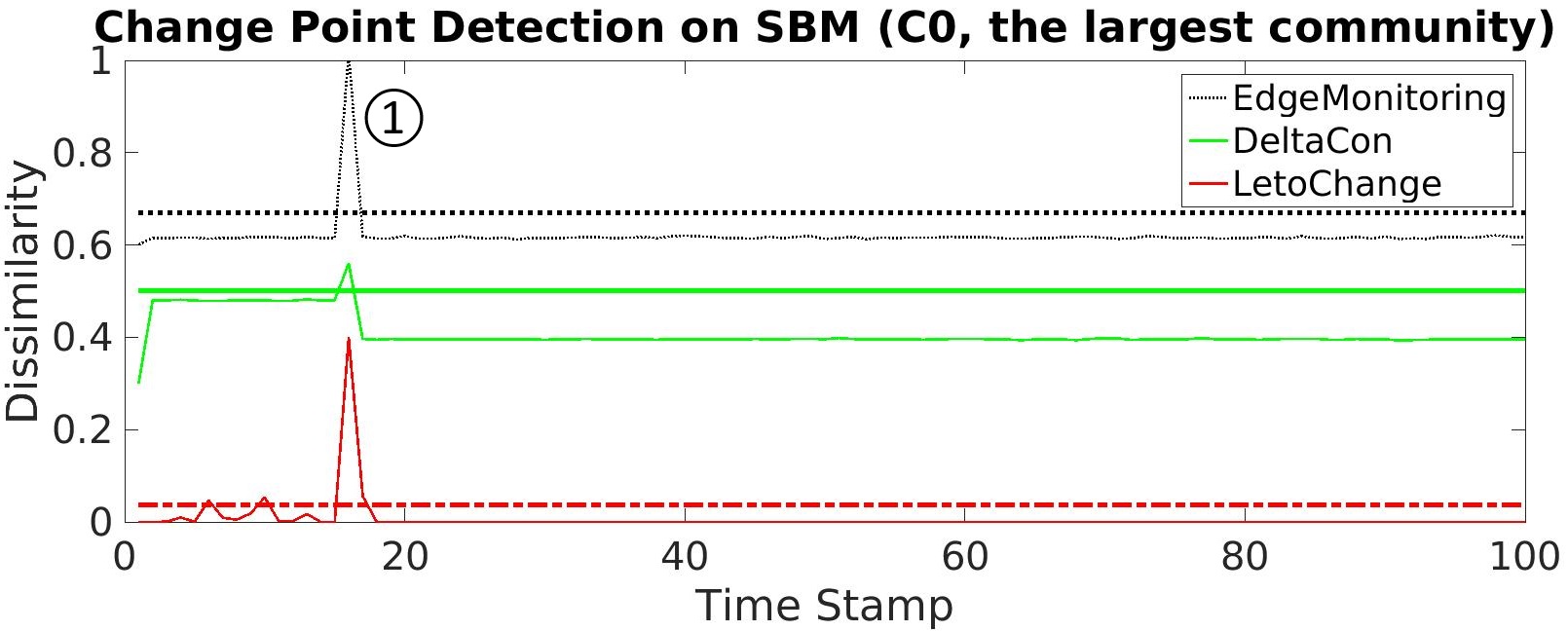}
\captionsetup{width=240pt}
\caption{\small{Change point detection on SBM in the largest community (300 nodes). All three algorithms are able to detect the local event \textcircled{1} at T=16.}}
\label{fig:SBM_C0}
\end{minipage}
\vspace{-0.2cm}
\end{figure}

\Cref{fig:SBM_C0} shows the change points detected on the largest community. 
All three algorithms detect the local change (\txtcircled{1} in \Cref{table:event}) that edge probability reduces at time stamp 16 in C0. 
It's interesting to note that EdgeMonitoring detects the local change \txtcircled{1} on the original network (\Cref{fig:SBMGlobalOriginal}). 
This is because the change occurs in the largest community (consisting of 30\% of all the nodes), 
and the node pairs in that community are more likely to be tracked. 
If we only run the algorithm on the original network, we would not be able to tell if the change corresponds to a global event or a local event. 
DeltaCon is more conservative in terms of flagging out change points: 
it not only misses the local event \txtcircled{1} in \Cref{fig:SBMGlobalOriginal}, 
but also misses the global event \txtcircled{2}. 
Yet it is still able to detect the local change in \Cref{fig:SBM_C0}. 
Similarly, running algorithms within community $c_6$, $c_7$ reveals local event \txtcircled{3}. This result is not included for brevity. 

\begin{figure}[!ht]
\begin{minipage}{240pt}
\subfloat[BTER Global (100 nodes, re-scaled and shifted for visualization). Local change \textcircled{1} is ignored by all three algorithms, while \textcircled{3} is captured by EdgeMonitoring (barely) and LetoChange.\label{fig:BTERGlobalOriginal}]{\includegraphics[width=240pt,height=110pt]{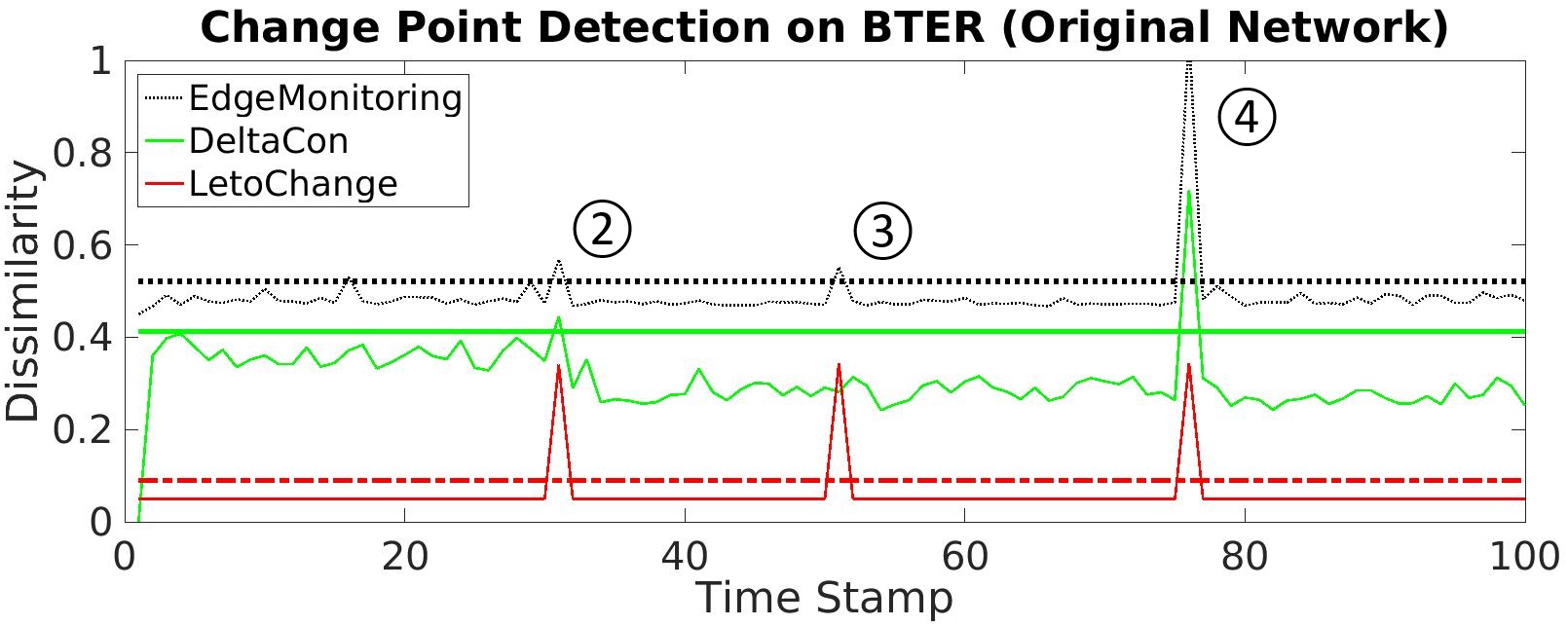}}\hfill
\subfloat[BTER Contract (5 hyper-nodes). EdgeMonitoring (black, top) and LetoChange (red, bottom) detect two global events, while DeltaCon (green, middle) detects only one global event \textcircled{4}.\label{fig:BTERGlobalContract}]{\includegraphics[width=240pt,height=110pt]{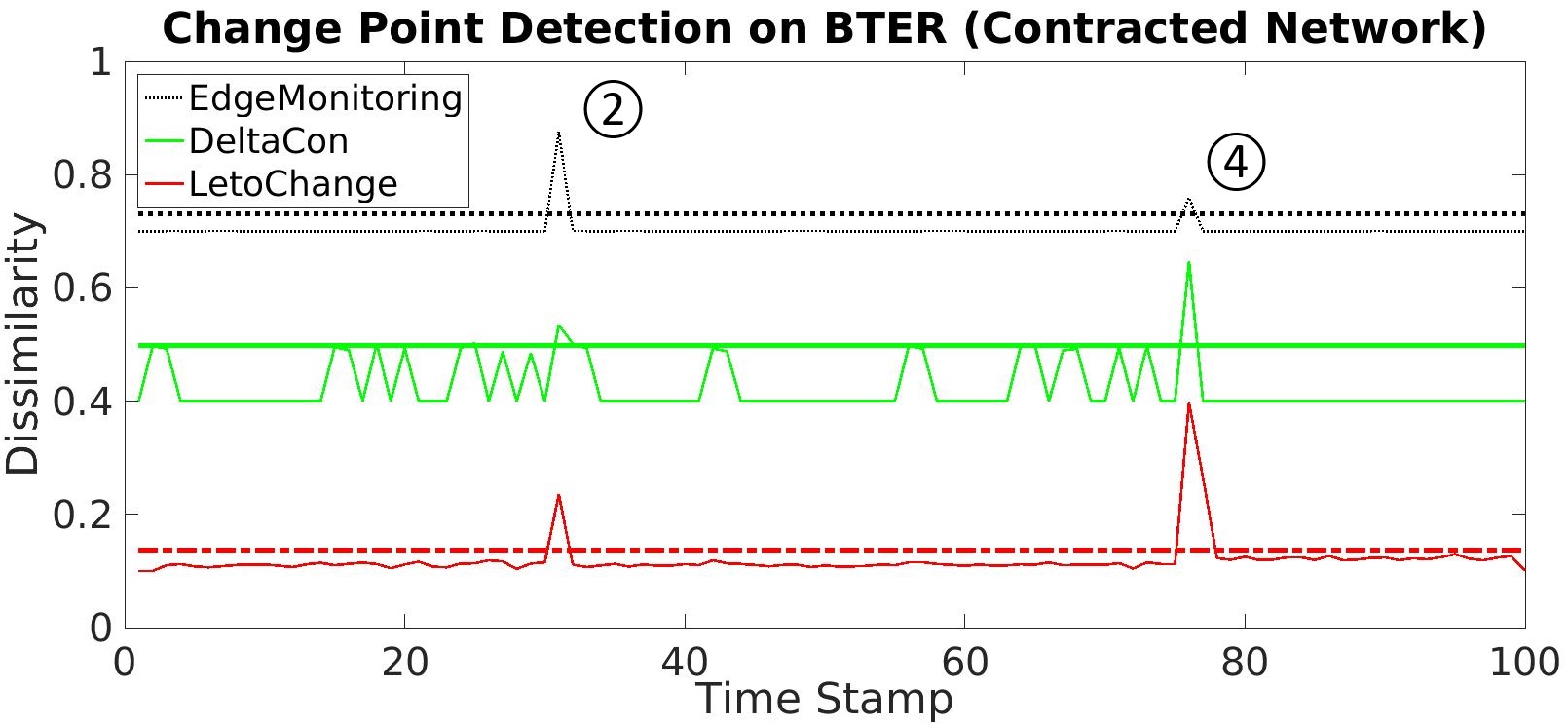}}\hfill
\vspace{-0.1cm}
\end{minipage}
\captionsetup{width=240pt}
\caption{\small{Change point detection on BTER at global level.}}
\label{fig:BTERGlobal}
\end{figure}

\begin{table*}[!t]
\small
\caption{\small{Time Efficiency Comparison on Different Networks}}
\label{table:efficiency}
\begin{threeparttable}
\begin{tabular}{|p{27pt}p{21pt}p{35pt}p{30pt}p{22pt}p{21pt}p{39pt}p{27pt}p{22pt}p{23pt}p{39pt}p{30pt}p{22pt}|}
\hline
 & \multicolumn{4}{|c|}{EdgeMonitoring\tnote{1}} & \multicolumn{4}{c|}{DeltaCon} & \multicolumn{4}{c|}{LetoChange}\\
\hline
Data & Orig. & Contr. (Speedup) & Comm. Avg & Total\tnote{2} & Orig. & Contr. (Speedup) & Comm. Avg & Total & Orig. & Contr. (Speedup) & Comm. Avg & Total \\
SBM & 85s & 1.5s (60X) & 3.7s & 31s & 500s & 0.6s\footnotesize{ \textbf{(830X)}} & 24s & 192s & DNF\tnote{3} & 2h ($>$60X) & N/A\tnote{4} & N/A\\
BTER & 2.5s & 1.5s (2X) & 1.7s & 10s & 5s & 0.1s (50X) & 1.0s & 5.1s & 51h & 2h (25X) & 13h & 67h\\
Trade & 22s & 0.4s (50X) & 0.5s & 3.4s & 42s & 1.3s (30X) & 1.2s & 8.5s & 150m & 5m (30X) & 9m & 60m\\
\hline
\end{tabular}
\begin{tablenotes}
\item[1]{EdgeMonitoring and DeltaCon are implemented in MATLAB and run on a commercial desktop, while LetoChange is implemented in Python and runs on a cluster with 28 cores. Each running time averaged over 5 runs.}
\item[2]{Total time $=$ community time $\times$ number of communities.}
\item[3]{The program runs for 120 hours, but still has no sign of finish, so we kill it.}
\item[4]{LetoChange takes 110 hours to obtain the result in~\Cref{fig:SBM_C0}.}
\end{tablenotes}
\end{threeparttable}
\end{table*}

\begin{figure}[!ht]
\begin{minipage}{240pt}
\subfloat[BTER network at T=75, before the global event \textcircled{4}\label{fig:BTER_visual_G1}]{\includegraphics[width=240pt,height=150pt]{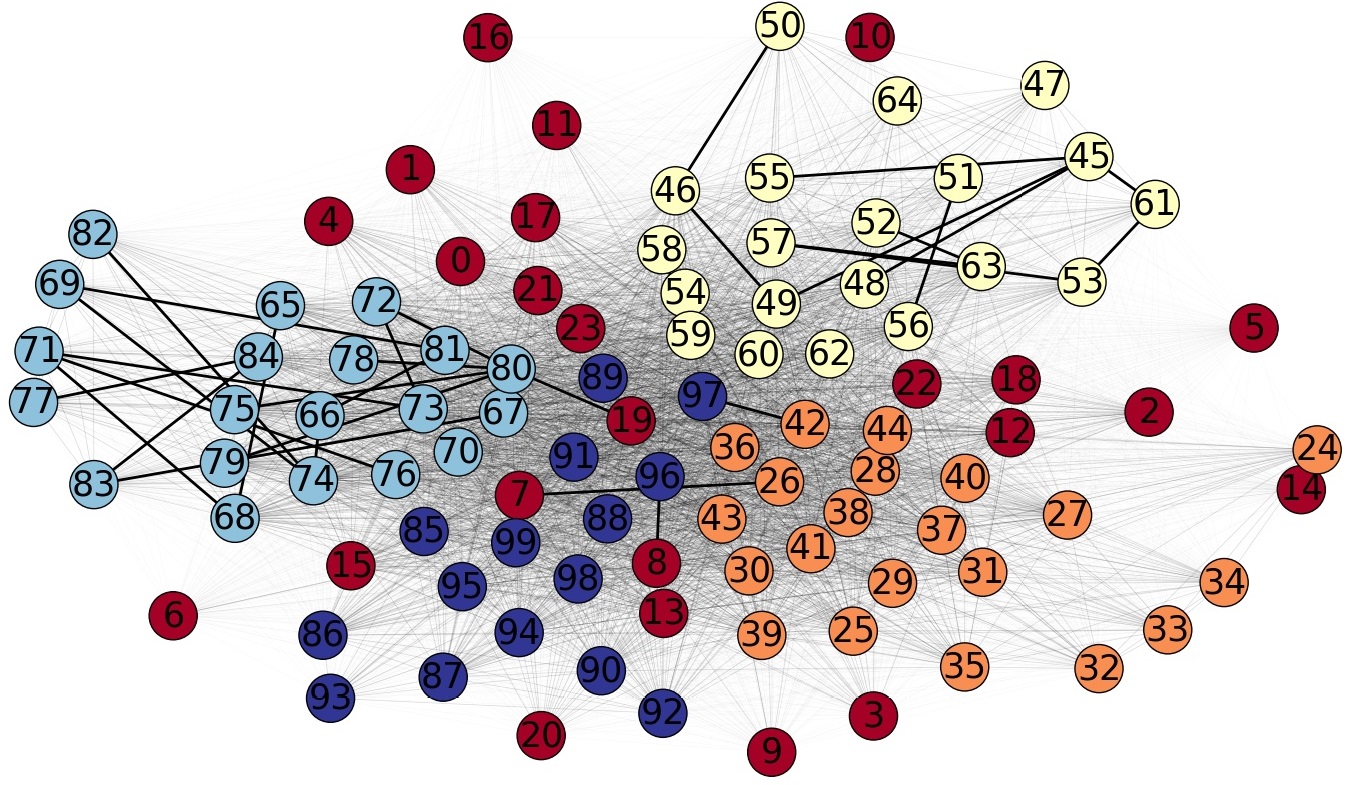}}\hfill
\subfloat[BTER network at T=76, after the global event \textcircled{4}\label{fig:BTER_visual_G2}]{\includegraphics[width=240pt,height=150pt]{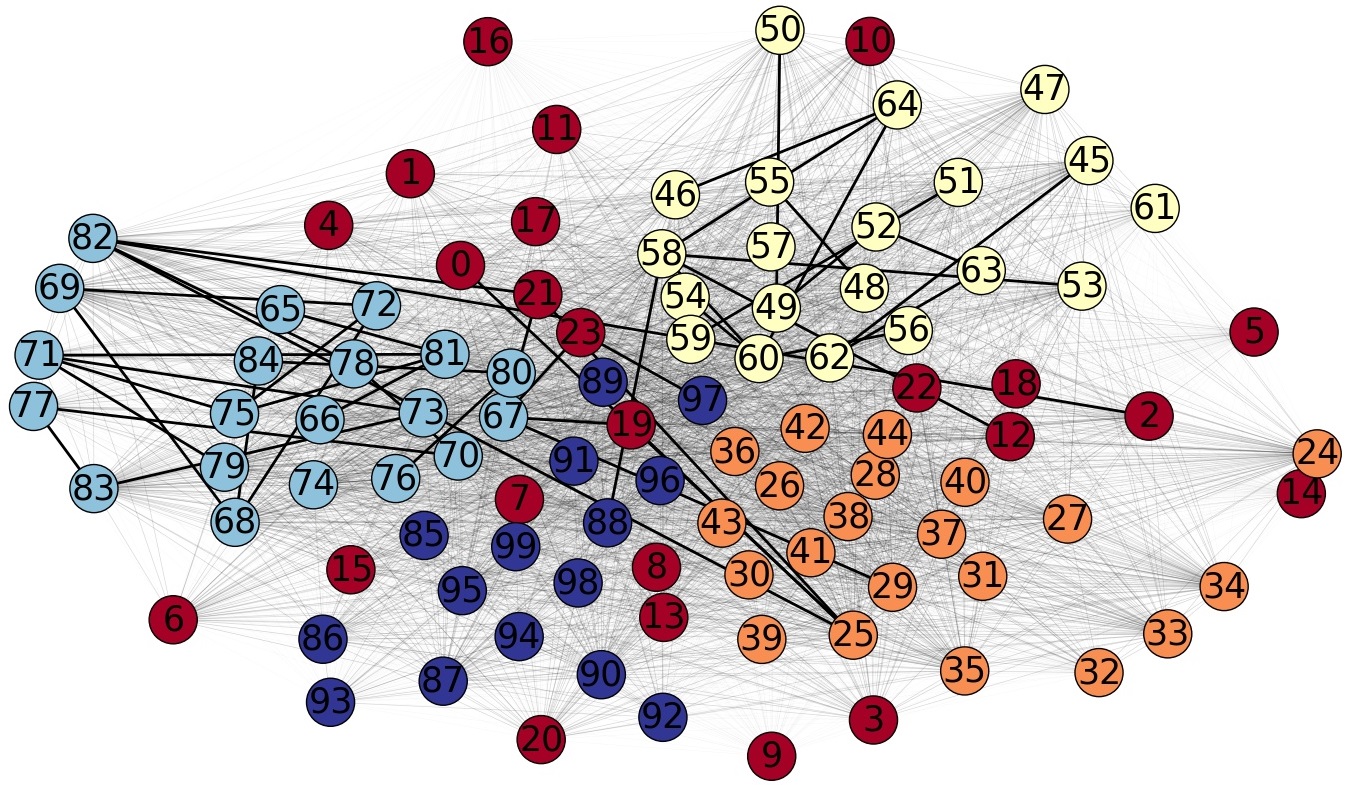}}\hfill
\vspace{-0.1cm}
\end{minipage}
\captionsetup{width=240pt}
\caption{\small{Visualization of BTER network before and after the global event \textcircled{4}. Edge width and transparency reflect interaction strength, and color represents community membership. Inter-community interaction is enhanced after the global event \textcircled{4}, and the red community in particular has stronger interaction with all other communities after \textcircled{4}.}}
\label{fig:BTER_visual}
\vspace{-0.3cm}
\end{figure}
\begin{figure}[!ht]
\begin{minipage}{240pt}
\subfloat[Contracted BTER network at T=75, before the global event \textcircled{4}\label{fig:BTER_visual_C1}]{\includegraphics[width=100pt,height=70pt]{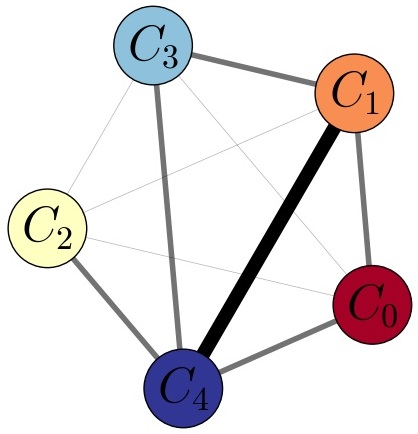}}\hspace{1.2cm}
\subfloat[Contracted BTER network at T=76, after the global event \textcircled{4}\label{fig:BTER_visual_C2}]{\includegraphics[width=100pt,height=70pt]{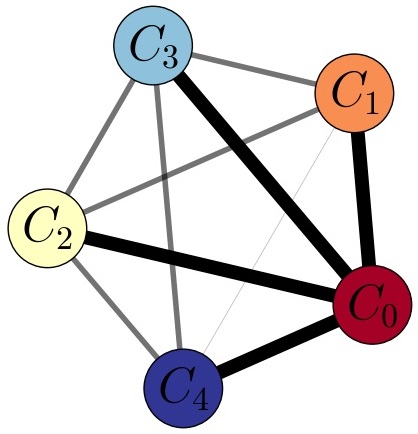}}\hfill
\vspace{-0.1cm}
\end{minipage}
\captionsetup{width=240pt}
\caption{\small{Visualization of contracted BTER network before and after the global event \textcircled{4}. The observation that the red community has stronger interaction with others in~\Cref{fig:BTER_visual} is nicely captured by the contracted network.}}
\label{fig:BTER_visual_contract}
\vspace{-0.5cm}
\end{figure}

\Cref{fig:BTERGlobal} shows the change points detected on the BTER network. 
We see that all three algorithms capture the global events \txtcircled{2}, \txtcircled{4} on the original network.
EdgeMonitoring and LetoChange in addition capture the local event \txtcircled{3}, 
which corresponds to edge probability increase in the two smallest communities (\Cref{table:event}). 
Although both \txtcircled{1} and \txtcircled{3} are local events, 
the largest community consists of 23\% of all the nodes while the two smallest communities together consist of 35\% of all the nodes. 
Intuitively, the latter should have more impact on the global network than the former. 
And from the algorithmic perspective, the node pairs of the latter have higher chance to be tracked than the former. 
\Cref{fig:BTERGlobalContract} shows that after contraction, all three algorithms only capture the global events. 
Running the algorithms within each community detects the local change and we do not include the result for brevity. 
The NDCG scores for EdgeMonitoring, DeltaCon and LetoChange are $0.832$, $0.666$ and $0.832$ respectively. 
The snapshots before and after the global event \txtcircled{4} are visualized in~\Cref{fig:BTER_visual,fig:BTER_visual_contract}. 
We can see from~\Cref{fig:BTER_visual} that the inter-community interaction between the red community and the others is enhanced after the event, 
which is also revealed in the contracted network in~\Cref{fig:BTER_visual_contract}. 

The two synthetic experiments above show that superimposing our framework on top of the aforementioned change point detection algorithms can indeed distinguish global and local change, 
and hence gives us finer granularity knowledge of the evolution of a dynamic network. 

\subsubsection{Real World Network}
\begin{figure}[!ht]
\begin{minipage}{240pt}
\subfloat[Original Network, both EdgeMonitoring (black star, top) and LetoChange (red diamond, bottom) 
detect the year 2009 as a change point, at which the international trade volume drops significantly 
(background bar); DeltaCon detects the year 2012 as abnormal. 
(re-scaled and shifted for visualization, explainable outliers are blue-boxed)
\label{fig:TradeGlobalOriginal}]{\includegraphics[width=240pt,height=140pt]{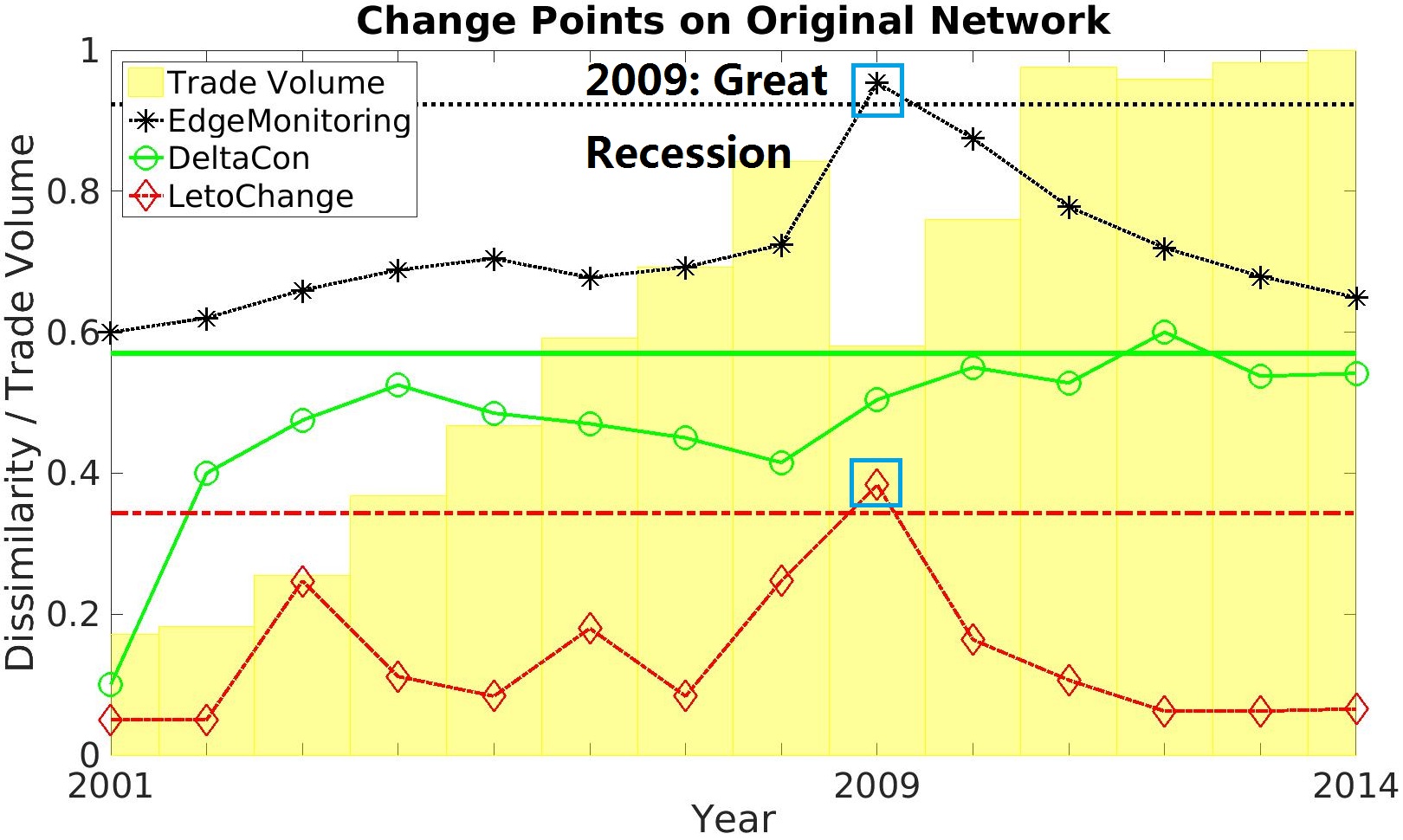}}\hfill
\subfloat[Contracted Network, all three algorithms detect the year 2009 as a change point.\label{fig:TradeGlobalContract}]{\includegraphics[width=240pt,height=140pt]{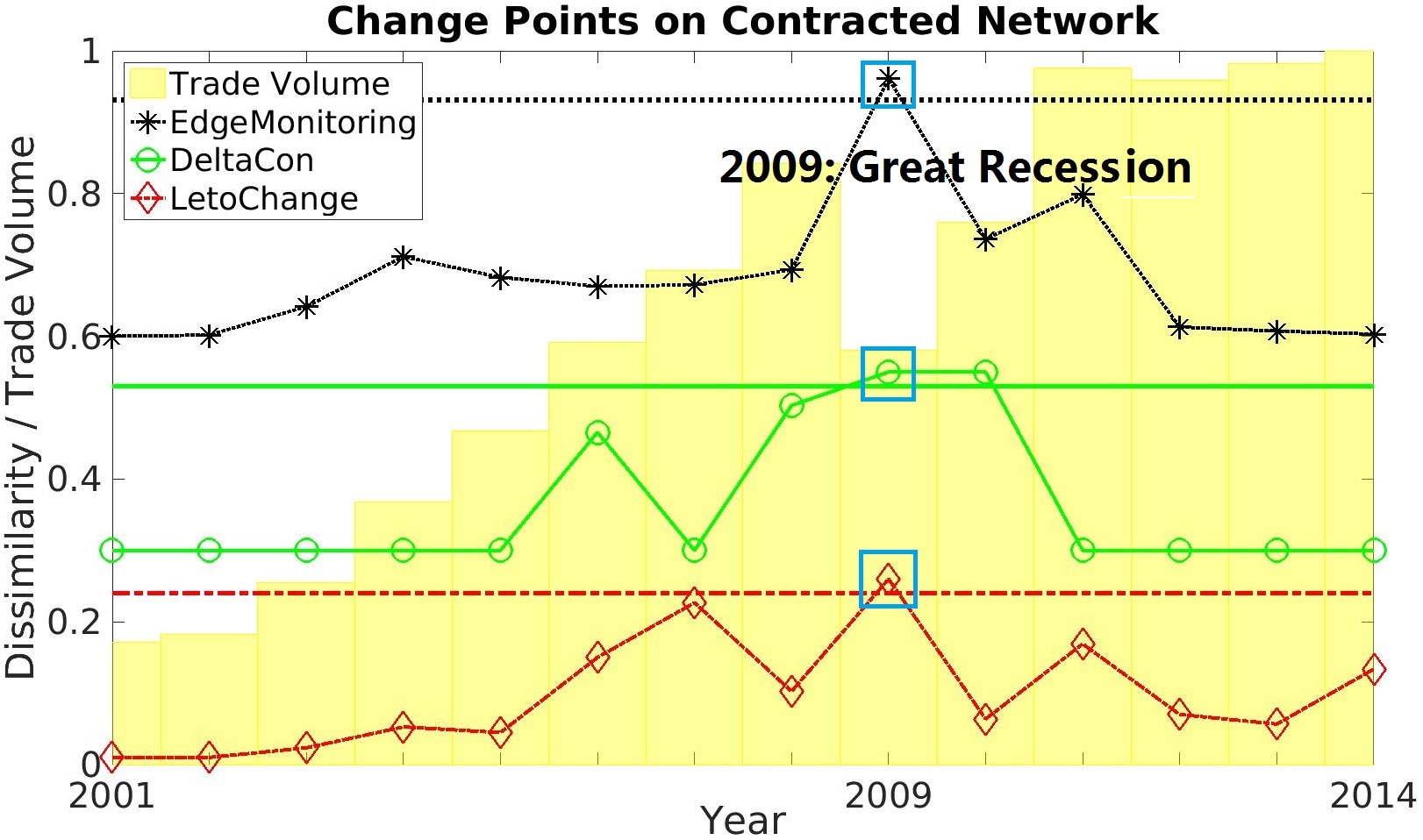}}\hfill
\vspace{-0.1cm}
\end{minipage}
\captionsetup{width=240pt}
\caption{\small{Global change point detected on the international trade network, using both the original network and the contracted network. The Great Recession (global event) in the year 2009 is flagged as abnormal on both the networks. The NDCG scores of the three algorithms on the two networks are $0.960,0.654,0.909$ respectively, indicating good approximation of the contracted network.}}
\label{fig:TradeGlobal}
\vspace{-0.5cm}
\end{figure}

\Cref{fig:TradeGlobal} shows the result of change point detection on the original network and the contracted network using the international trade network. 
It can be seen that the majority algorithms on both the networks reveal the year 2009 as the most significant outlier. 
The year 2009 is a year immediately after the global financial crisis, and the term ``The Great Recession'' is applied to the global recession which started in that year.\footnote{\footnotesize{http://www.cbpp.org/research/economy/chart-book-the-legacy-of-the-great-recession}}
It is pretty clear that the global trade volume drops significantly during that year. 
The NDCG scores for EdgeMonitoring, LetoChange and DeltaCon are $0.960$, $0.654$ and $0.909$ respectively, 
which suggests great approximation for EdgeMonitoring and LetoChange, and good approximation for DeltaCon. 
The speedup of the framework is reported in~\Cref{table:efficiency}.

\begin{figure}[!htb]
\begin{minipage}{240pt}
\subfloat[Global network.\label{fig:TradeGlobalAgain}]{\includegraphics[width=240pt,height=120pt]{./Pics/GlobalNetworkThreeAlgos_Annotated.jpg}}\hfill
\subfloat[APEC sub-network. Both the years 2009 and 2010 are flagged out as outliers. 
We did not find a good explanation for LetoChange's outlier at 2006. 
NDCG scores to the global network are 0.913, 0.827, 0.975 respectively.
\label{fig:APEC}]{\includegraphics[width=240pt,height=130pt]{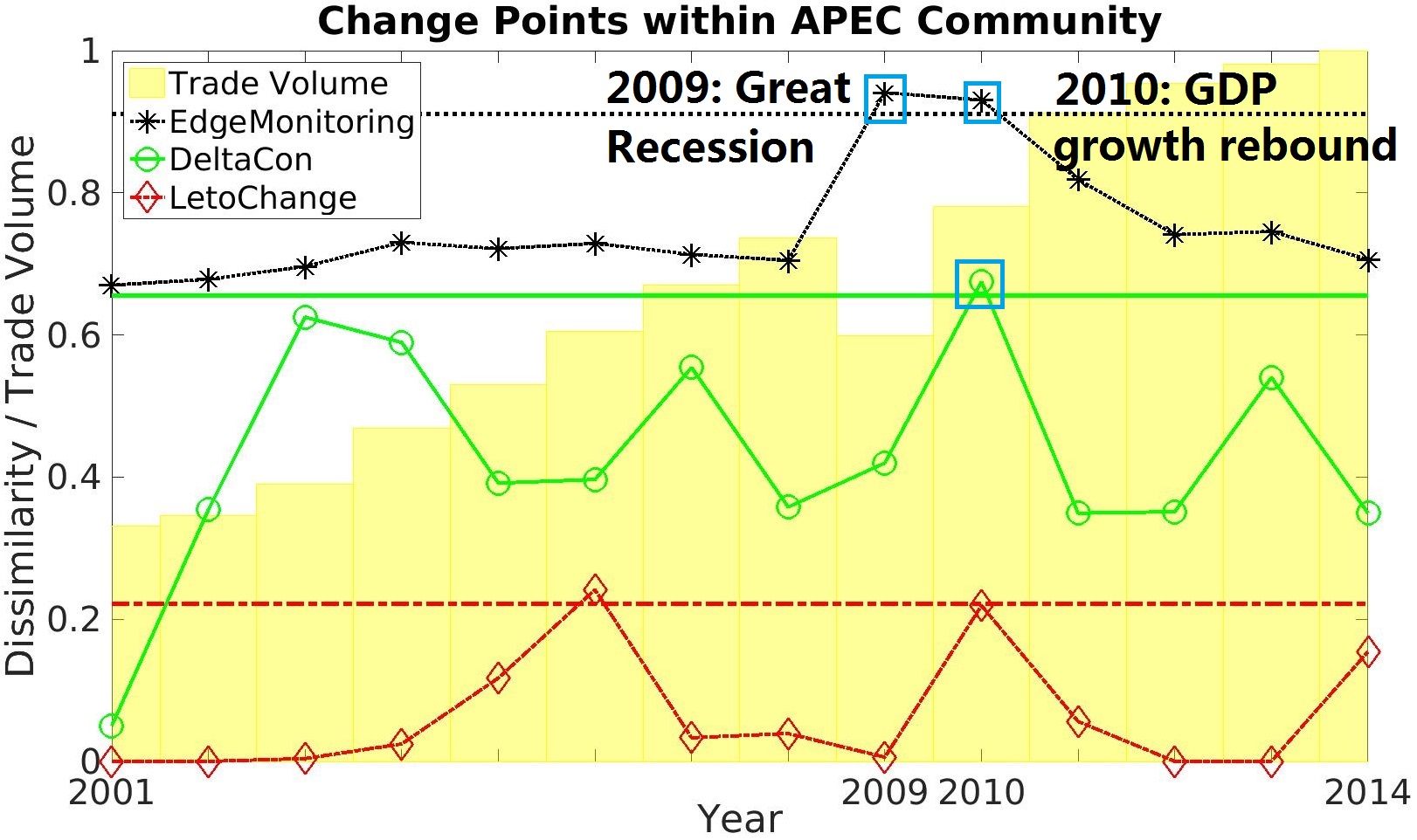}}\hfill
\subfloat[OIC sub-network. The evolution pattern is significantly different from that of the global network in~\Cref{fig:TradeGlobalAgain}. NDCG scores to the global network are 0.826, 0.714, 0.906 respectively.\label{fig:OIC}]{\includegraphics[width=240pt,height=130pt]{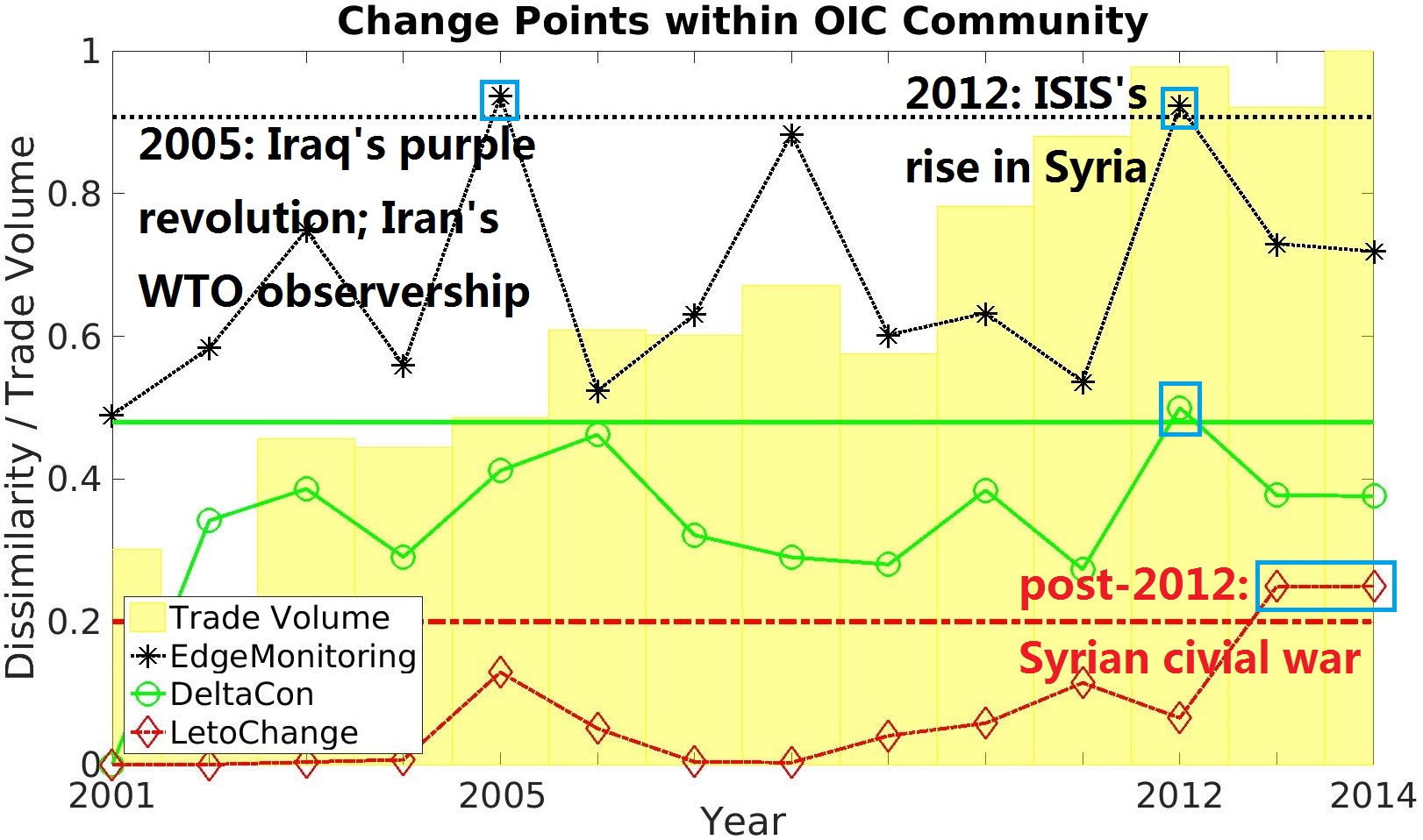}}\hfill
\vspace{-0.1cm}
\end{minipage}
\captionsetup{width=240pt}
\caption{\small{Comparison of global vs local. APEC community's evolution resembles that of the global network in both trade volume and outlier ranking due to its major role (heaviest edges) in the global network. 
OIC community evolves differently from the global network, and changes associated to OIC community is missing in the global network.}}
\label{fig:TradeGlobalLocalComparison}
\vspace{-0.6cm}
\end{figure}

\begin{figure}[!htb]
\begin{minipage}{240pt}
\subfloat[APEC trade network in 2008.\label{fig:NetworkAPEC2008}]{\includegraphics[width=240pt,height=120pt]{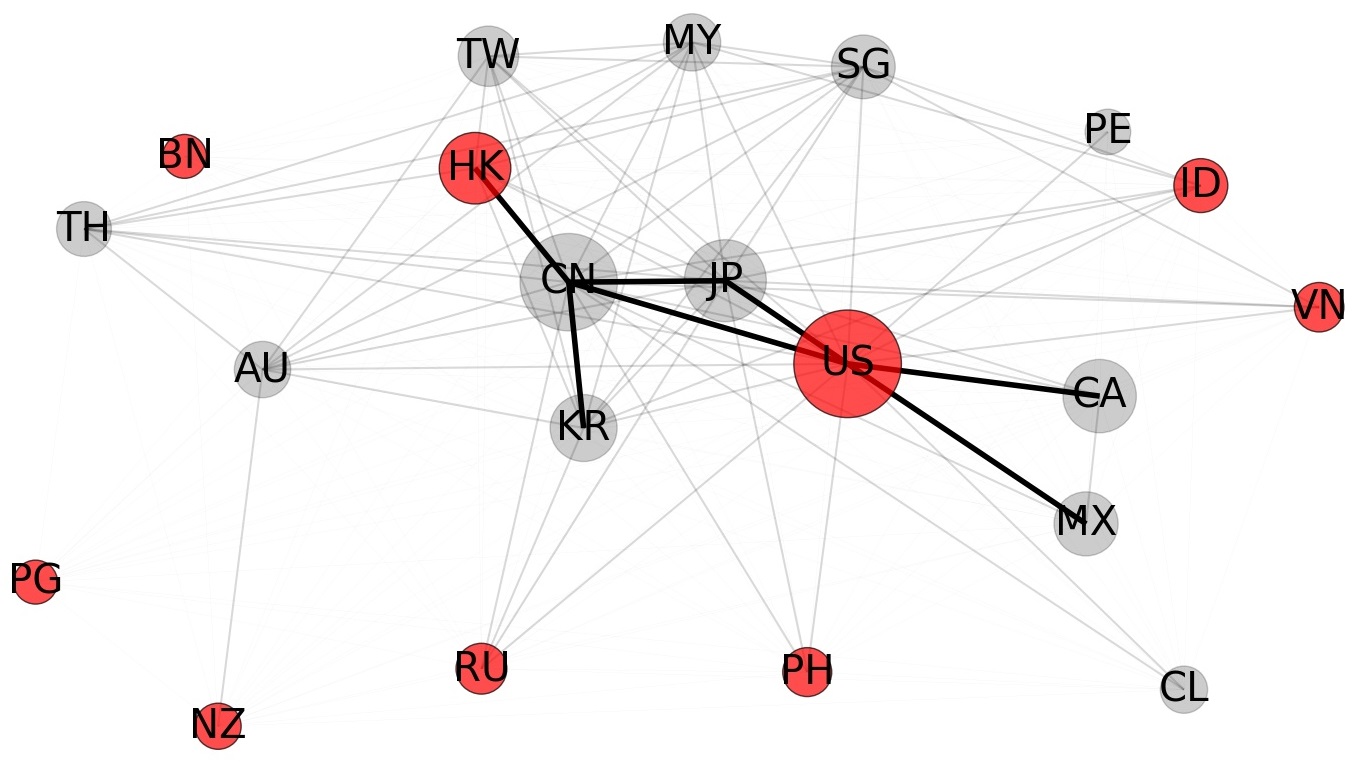}}\hfill
\subfloat[APEC trade network in 2009. International trade volume drops significantly (fewer heavy edges), 
and several economics flip between surplus and deficit (within the region).\label{fig:NetworkAPEC2009}]{\includegraphics[width=240pt,height=130pt]{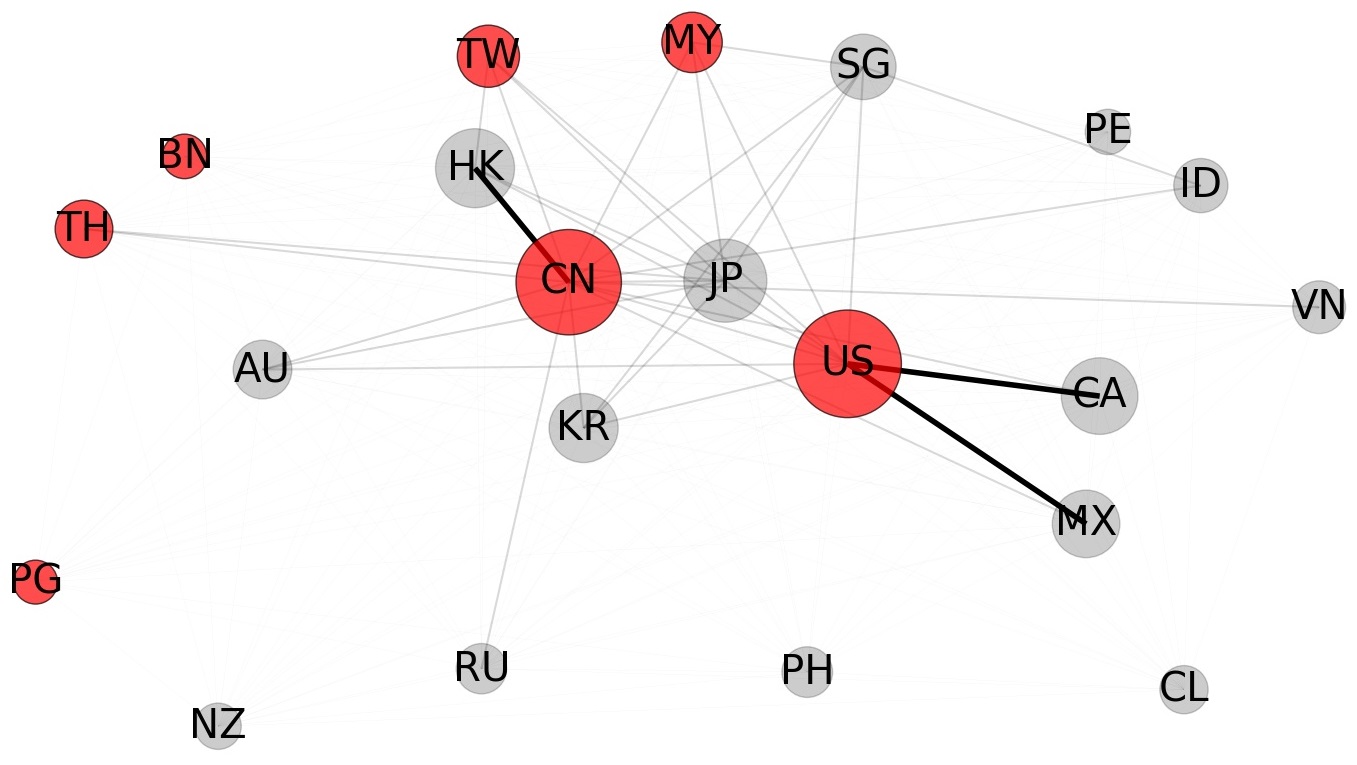}}\hfill
\subfloat[APEC trade network in 2010. More and heavier edges emerge, sign of recovery.\label{fig:NetworkAPEC2010}]{\includegraphics[width=240pt,height=130pt]{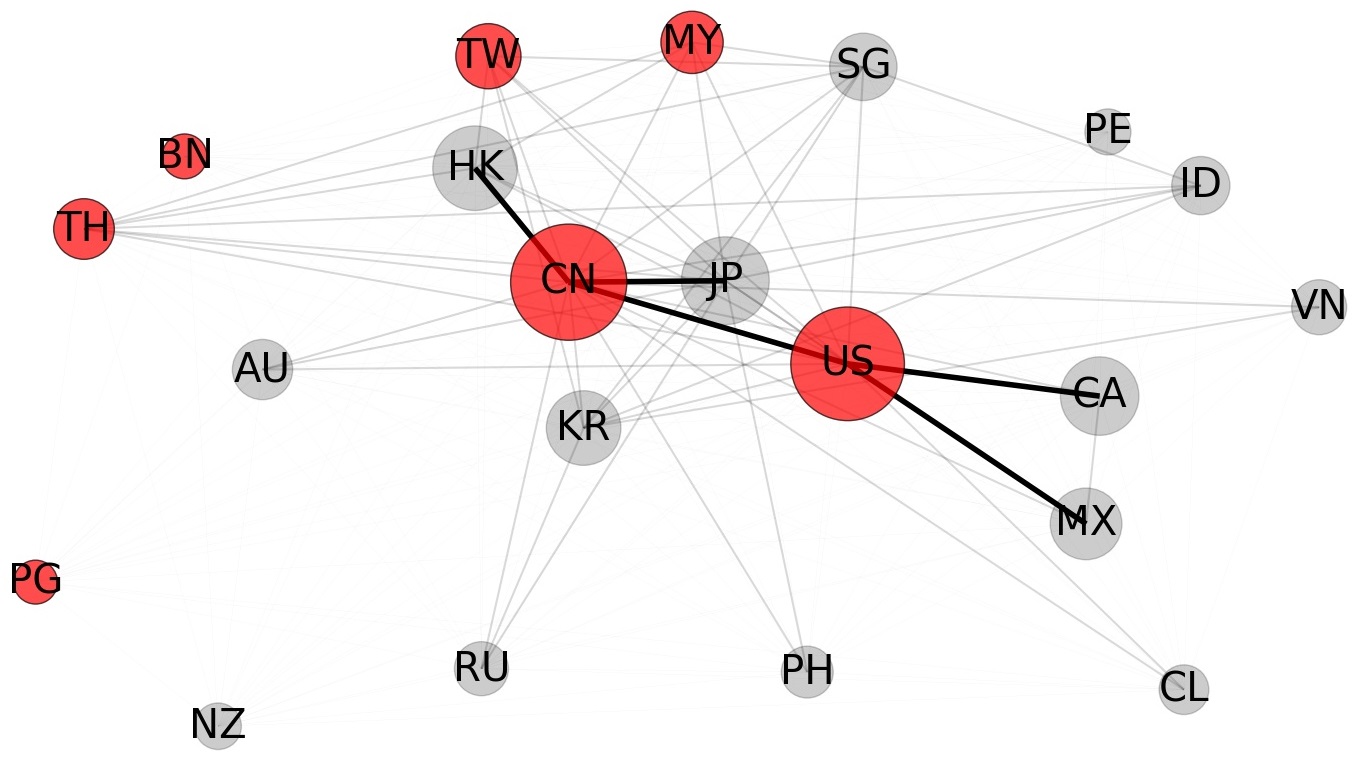}}\hfill
\vspace{-0.1cm}
\end{minipage}
\captionsetup{width=240pt}
\caption{\small{Visualization of APEC trade network (weighted, undirected). 
Node size is proportional to the economic size, and red nodes have trade deficit while grey nodes have trade surplus. 
Edge width and transparency reflect bilateral trade volume (export+import).}}
\label{fig:APECNetwork}
\vspace{-0.5cm}
\end{figure}

\Cref{fig:TradeGlobalLocalComparison} compares global change point detection and local change point detection. 
\Cref{fig:TradeGlobalAgain} is the result of running the detection algorithms on the original network. 
\Cref{fig:APEC} is the result of running the detection on the APEC community (Asia-Pacific region). 
We see that the two (sub-)networks have similar outliers: the 2009 recession is detected on both the networks. 
This is not surprising since this APEC community contributes $44\%$ of the world trade,\footnote{\footnotesize{https://ustr.gov/trade-agreements/other-initiatives/asia-pacific-economic-cooperation-apec/us-apec-trade-facts}} 
which can also be observed from the similarity of the two background trade volume bar charts. 
Here we see the impact of a ``big'', in terms of edge weight, community to the global network. 
Two detection algorithms also flag out the year 2010 as a local change point, 
which could be explained by the GDP growth rebound of the major economies in this region.\footnote{\footnotesize{http://data.worldbank.org/indicator/NY.GDP.MKTP.KD.ZG}} 
\Cref{fig:APECNetwork} visualizes the APEC trade network. We can see that the bilateral trade volume drops significantly post crisis, 
and several economics flip from trade surplus to trade deficit within the region. 
Comparing \Cref{fig:NetworkAPEC2010} and \Cref{fig:NetworkAPEC2009}, we see trade volume jumps among several economics, which is a sign of recovery.

\Cref{fig:OIC} shows the result of local change point detection on the Middle East-North Africa community. 
Its evolution pattern (the trade volume bar chart, as well as the outlier score ranking distribution) is clearly different from that of the global network. 
This can also be expected given its relatively small scale economy size and relatively homogeneous economic structures. 
The year 2005 is flagged as an outlier by EdgeMonitoring, which coincides with Iraq's purple revolution.
\footnote{\footnotesize{https://en.wikipedia.org/wiki/Colour\_revolution}} 
The year 2012 is flagged as an outlier by both EdgeMonitoring and DeltaCon, which coincides with ISIS's rise in Syria.\footnote{\footnotesize{https://en.wikipedia.org/wiki/Islamic\_State\_of\_Iraq\_and\_the\_Levant}} 
LetoChange flags the years 2013 and 2014 as outliers, 
we conjecture that might be due to the escalation of the situation with increasing international involvement.  

Both change points and outlier ranking distribution reflect network evolution pattern. 
From \Cref{fig:TradeGlobalLocalComparison} we see that while the evolution pattern of a major community might approximate that of the global network, 
the evolution patterns of other communities differ from that of the global network. 
We will miss the change points or events associated with a particular community if we only run the detection algorithm on the global network. 

\section{Conclusion}
In this paper, we study hierarchical change point detection on dynamic social networks. 
We distinguish the intra-community evolution and the inter-community evolution. 
Our framework detects global change points on the inter-community network, 
and local change points on the intra-community networks. 
This framework is compatible with several state-of-the-art change point detection algorithms. 
Extensive empirical evaluation on several networks (both synthetic and real world) show this framework has not only quality advantages but also significant computational benefits.

\begin{acks}
The authors would like to thank Leto Peel for providing his code base, and Wenlei Bao and and Congrong Guan for helping collect the international trade data. 
The authors would also like to thank the anonymous referees for their valuable comments. 
This work is supported in part by NSF grant DMS-1418265, IIS-1550302 and IIS-1629548. Any opinions, findings, and conclusions or recommendations expressed in this material are those of the authors and do not necessarily reflect the views of the National Science Foundation.
\end{acks}
\bibliographystyle{ACM-Reference-Format}
\bibliography{myRef} 
\balance

\end{document}